\documentclass[12pt]{article}
\usepackage{graphicx}
\begin{document}

\vbox{\vspace{5ex}}

\begin{center}
{\Large \bf Can you do quantum mechanics without Einstein?}

\vspace{2ex}

Y. S. Kim \footnote{electronic address: yskim@physics.umd.edu}\\
Department of Physics, University of Maryland,\\
College Park, Maryland 20742, U.S.A.\\

\vspace{2ex}

Marilyn E. Noz \footnote{electronic address: nozm01@med.nyu.edu}\\
Department of Radiology, New York University,\\
New York, New York 10016, U.S.A.\\

\end{center}

\vspace{2ex}
\begin{abstract}
The present form of quantum mechanics is based on the Copenhagen
school of interpretation.  Einstein did not belong to the Copenhagen
school, because he did not believe in probabilistic interpretation of
fundamental physical laws.  This is the reason why we are still
debating whether there is a more deterministic theory.
One cause of this separation between Einstein and the Copenhagen
school could have been that the Copenhagen physicists thoroughly
ignored Einstein's main concern: the principle of relativity.
Paul A. M. Dirac was the first one to realize this problem.
Indeed, from 1927 to 1963, Paul A. M. Dirac published at least
four papers to study the problem of making the uncertainty relation
consistent with Einstein's Lorentz covariance.   It is interesting
to combine those papers by Dirac to make the uncertainty relation
consistent with relativity.  It is shown that the mathematics of
two coupled oscillators enables us to carry out this job.  We are
then led to the question of whether the concept of localized
probability distribution is consistent with Lorentz covariance.

\end{abstract}

\newpage

\section{Introduction}\label{intro}

\begin{figure}[thb]
\centerline{\includegraphics[scale=0.5]{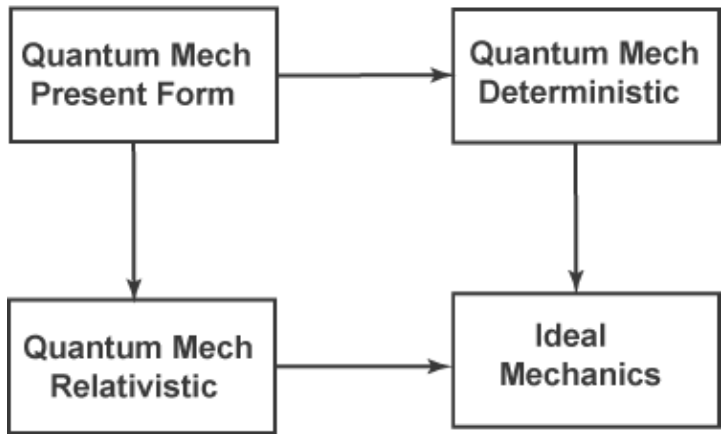}}
\vspace{5mm}
\caption{Toward ideal mechanics.  The ideal mechanics should be both
deterministic and relativistic.  Einstein had enough reason to complain
about the lack of Lorentz covariance in the Copenhagen interpretation,
but he was reticent about it.  Instead, Einstein concentrated his
opposition on the probabilistic interpretation.  Why?}\label{ideal}
\end{figure}

Einstein was against the Copenhagen interpretation of quantum
mechanics.  Why was he so against it?  The present form of quantum
mechanics is regarded as unsatisfactory because of its probabilistic
interpretation.  At the same time, it is unsatisfactory because
it does not appear to be Lorentz-covariant.  We still do not
know how the hydrogen atom appears to a moving observer.  Indeed,
we have to go through two-track routes to reach the ideal
mechanics, as illustrated in Fig.~\ref{ideal}.

While relativity was Einstein's main domain of interest, why did he
not complain about the lack of Lorentz covariance?  It is possible
that Einstein was too modest to mention relativity, and instead
concentrated his complaint against its probabilistic interpretation.
It is also possible that Einstein did not want to sent his most
valuable physics asset to a battle ground.  We cannot find a definite
answer to this question, but it is gratifying to note that the
present authors are not the first ones to question whether the
Copenhagen school of thought is consistent with the concept of
relativity.

Paul A. M. Dirac was never completely happy with the Copenhagen
interpretation of quantum mechanics, but he thought it was a
necessary temporary step.  In that case, he thought we should
examine whether quantum mechanics is consistent with special
relativity.

\begin{figure}[thb]
\centerline{\includegraphics[scale=0.6]{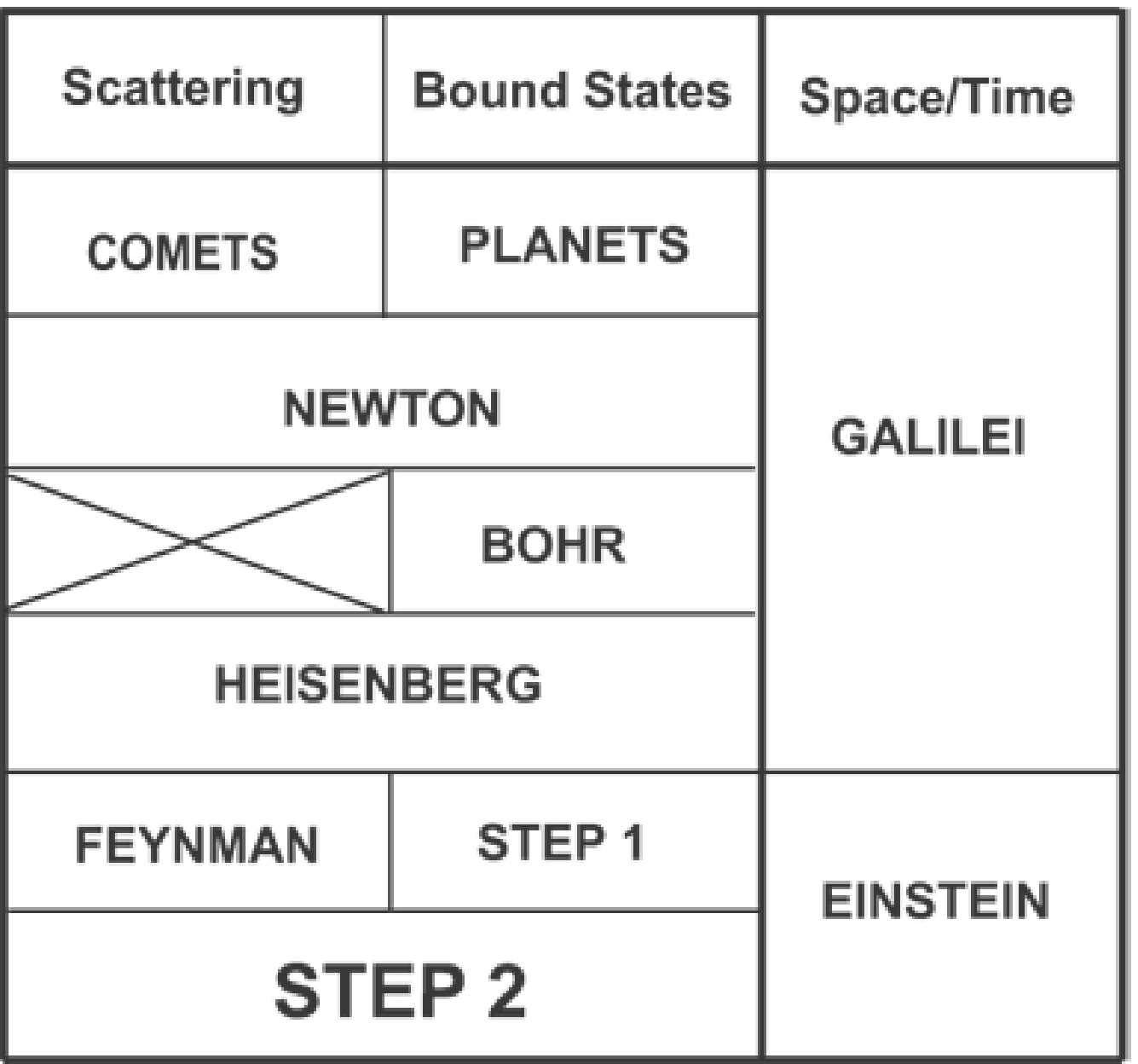}}
\vspace{5mm}
\caption{History of dynamical and kinematical developments.  It is
important to note that mankind's unified understanding of scattering
and bound states has been very brief.  It is therefore not unusual to
expect that separate theoretical models be developed for scattering and
for bound states.  The successes and limitations of the Feynman diagram
are well known.  If we cannot build a covariant quantum mechanics,
it is worthwhile to see whether we can construct a relativistic theory
of bound states to supplement quantum field theory, as Step 1 before
attempting to construct a Lorentz-covariant theory applicable to both
in Step 2.}\label{comet}
\end{figure}

As for combining quantum mechanics with special relativity, there
was a giant step of constructing the present form of quantum field
theory.  It leads to a Lorentz covariant S-matrix which enables us
to calculate scattering amplitudes using Feynman diagrams.  However,
we cannot solve bound-state problems or localized probability
distributions using Feynman diagrams~\cite{fkr71}.  We have to
construct a separate theoretical device to address this issue,
as illustrated in Fig.~\ref{comet}.

Dirac was never happy with the present form of field theory~\cite{dir70},
particularly with infinite quantities in its renormalization processes.
Furthermore, field theory never addresses the issue of localized
probability.  Indeed, Dirac concentrated his efforts in seeing
whether localized probability distribution is consistent with Lorentz
covariance.

In 1927~\cite{dir27}, Dirac
noted that there is a time-energy uncertainty relation without
time-like excitations.  He pointed out that this space-time
asymmetry causes a difficulty in combining quantum mechanics with
special relativity.

In 1945~\cite{dir45}, Dirac constructed four-dimensional
harmonic oscillator wave functions including the time variable.
His oscillator wave functions took normalizable Gaussian form,
but he did not attempt to give a physical interpretation to this
mathematical device.

In 1949~\cite{dir49}, Dirac emphasized that the task of building
a relativistic quantum mechanics is equivalent to constructing a
representation of the Poincar\'e group.  He then pointed out
difficulties in constructing such a representation.
He also introduced the light-cone coordinate system.

In 1963~\cite{dir63}, Dirac used two coupled oscillators
to construct a representation of the $O(3,2)$ deSitter group
which later became the basic mathematical base for two-photon
coherent states known as squeezed states of light~\cite{knp91}.

In this report, we combine all of these works by Dirac to
make the present form of uncertainty relations consistent
with special relativity.  Once this task is complete, we can
start examining whether the probability interpretation is
ultimately valid for quantum mechanics.

In Secs.~\ref{quantu} --~\ref{o32}, we examine each of the
above-mentioned papers of Dirac.  In Sec.~\ref{dirac1}, we
combine these four papers into one paper using the language
of coupled harmonic oscillators.

\section{Dirac's c-number Time-energy uncertainty relation}\label{quantu}

The time-energy uncertainty relation was known before 1927 from the
transition time and line broadening in atomic spectroscopy.  As soon as
Heisenberg formulated his uncertainty, Dirac considered whether this
uncertainty can be combined with the position momentum uncertainty to
form a Lorentz covariant uncertainty relation~\cite{dir27}.

He noted one major difficulty.  There are excitations along the
space-like longitudinal direction starting from the position-momentum
uncertainty, while there are no excitations along the time-like
direction. The time variable is a c-number.  How then can this
space-time asymmetry be made consistent with Lorentz covariance,
where space and time coordinate are mixed up for moving observers.

On the other hand, Dirac forgot to consider Heisenberg's uncertainty
relation is applicable to space separation variables.  For instance,
the Bohr radius measures the difference between the proton and electron.
Dirac never addressed the question of separation in time variable or
time interval even in his later papers.

\begin{figure}[thb]
\centerline{\includegraphics[scale=0.9]{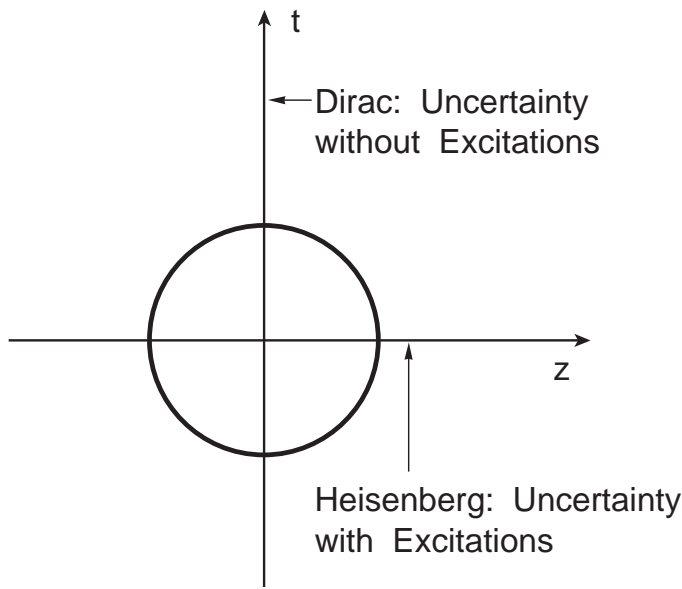}}
\vspace{5mm}
\caption{Space-time picture of quantum mechanics.  There
are quantum excitations along the space-like longitudinal direction, but
there are no excitations along the time-like direction.  The time-energy
relation is a c-number uncertainty relation.}\label{quantum}
\end{figure}

As for the space-time asymmetry, Dirac came back to this question in
his 1949 paper~\cite{dir49} where he discusses the ``instant form'' of
relativistic dynamics.  He talks about indirectly freezing the possibility
of three of the six parameters parameters of the Lorentz group, and thus
working only with three free parameters.  This idea was presented
earlier by Wigner~\cite{wig39,knp86} who observed that the internal
space-time symmetries of particles are dictated by his little groups
with three independent parameters.

\section{Dirac's four-dimensional oscillators}

During World War II, Dirac was looking into the possibility of
constructing representations of the Lorentz group using harmonic
oscillator wave functions~\cite{dir45}.  The Lorentz group is the
language of special relativity, and the present form of quantum
mechanics starts with harmonic oscillators.  Therefore, he was
interested in making quantum mechanics Lorentz-covariant by
constructing representations of the Lorentz group using harmonic
oscillators.

In his 1945 paper~\cite{dir45}, Dirac considers the Gaussian form
\begin{equation}\label{ground4}
\exp\left\{- {1 \over 2}\left(x^2 + y^2 + z^2 + t^2\right)\right\} .
\end{equation}
We note that this Gaussian form is in the $(x,~y,~z,~t)$
coordinate variables.  Thus, if we consider a Lorentz boost along the
$z$ direction, we can drop the $x$ and $y$ variables, and write the
above equation as
\begin{equation}\label{ground}
\exp\left\{- {1 \over 2}\left(z^2 + t^2\right)\right\} .
\end{equation}
This is a strange expression for those who believe in Lorentz invariance
where $\left(z^2 - t^2\right)$ is an invariant quantity.

On the other hand, this expression is consistent with his earlier papers
on the time-energy uncertainty relation~\cite{dir27}.  In those papers,
Dirac observed that there is a time-energy uncertainty relation, while
there are no excitations along the time axis.

Let us look at Fig.~\ref{quantum} carefully.  This figure is a pictorial
representation of Dirac's Eq.(\ref{ground}),  with localization in both
space and time coordinates.  Then Dirac's fundamental question would be
how to make this figure covariant?  This is where Dirac stops.  However,
this is not the end of the Dirac story.

\section{Dirac's light-cone coordinate system}\label{lightcone}

In 1949, the Reviews of Modern Physics published a special issue to
celebrate Einstein's 70th birthday.  This issue contains Dirac paper
entitled ``Forms of Relativistic Dynamics''~\cite{dir49}.
In this paper, he introduced his light-cone coordinate system,
in which a Lorentz boost becomes a squeeze transformation.

When the system is boosted along the $z$ direction, the transformation
takes the form
\begin{equation}\label{boostm}
\pmatrix{z' \cr t'} = \pmatrix{\cosh(\eta/2) & \sinh(\eta/2) \cr
\sinh(\eta/2) & \cosh(\eta/2) } \pmatrix{z \cr t} .
\end{equation}

\begin{figure}[thb]
\centerline{\includegraphics[scale=0.8]{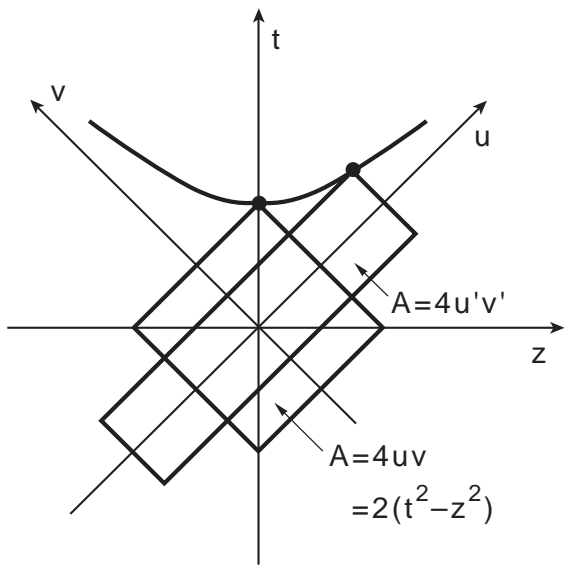}}
\vspace{2mm}
\caption{Lorentz boost in the light-cone coordinate system.  The boost
traces a point along the hyperbola.  The boost also squeezes the
square into a rectangle.}\label{licone}
\end{figure}

This is not a rotation, and people still feel strange about this
form of transformation.  In 1949~\cite{dir49}, Dirac introduced his
light-cone
variables defined as~\cite{dir49}
\begin{equation}\label{lcvari}
u = (z + t)/\sqrt{2} , \qquad v = (z - t)/\sqrt{2} ,
\end{equation}
the boost transformation of Eq.(\ref{boostm}) takes the form
\begin{equation}\label{lorensq}
u' = e^{\eta/2 } u , \qquad v' = e^{-\eta/2 } v .
\end{equation}
The $u$ variable becomes expanded while the $v$ variable becomes
contracted, as is illustrated in Fig.~\ref{licone}.  Their product
\begin{equation}
uv = {1 \over 2}(z + t)(z - t) = {1 \over 2}\left(z^2 - t^2\right)
\end{equation}
remains invariant.  In Dirac's picture, the Lorentz boost is a
squeeze transformation.

\begin{figure}[thb]
\centerline{\includegraphics[scale=0.5]{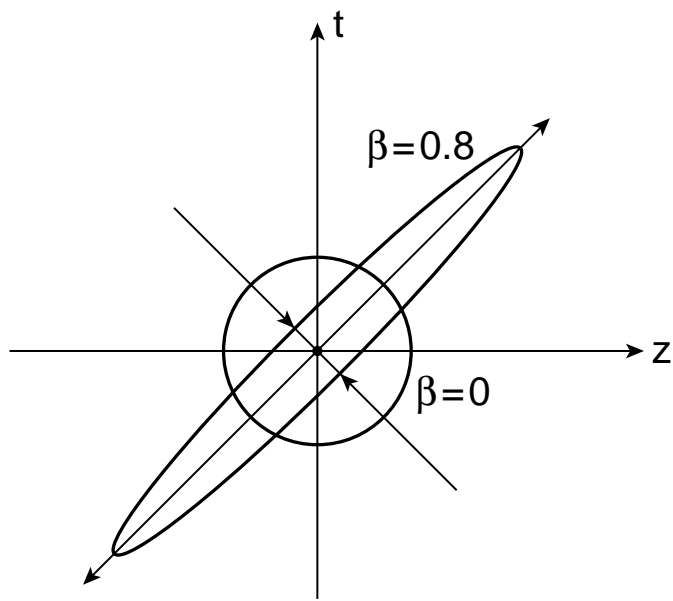}}
\caption{Effect of the Lorentz boost on the space-time
wave function.  The circular space-time distribution in the rest frame
becomes Lorentz-squeezed to become an elliptic
distribution.}\label{ellipse}
\end{figure}

If we combine Fig.~\ref{quantum} and Fig.~\ref{licone}, then we end up
with Fig.~\ref{ellipse}.
In mathematical formula, this transformation changes the Gaussian form
of Eq.(\ref{ground}) into
\begin{equation}\label{eta}
\psi_{\eta }(z,t) = \left({1 \over \pi }\right)^{1/2}
\exp\left\{-{1\over 4}\left[e^{-\eta }(z + t)^{2} +
e^{\eta}(z - t)^{2}\right]\right\} .
\end{equation}
This formula together with Fig.~\ref{ellipse} is known to describe
all essential high-energy features observed in high-energy
laboratories~\cite{fey69,kn77par,knp86,kn05job}.

Indeed, this elliptic deformation explains one of the most controversial
issues in high-energy physics.  Hadrons are known to be bound states of
quarks.  Its bound-state quantum mechanics is assumed to be the same as
that of the hydrogen atom.  The question is how the hadron would look
to an observer on a train.  If the train moves with a speed close to that
of light, the hadron appears like a collection of partons, according to
Feynman~\cite{fey69}.  Feynman's partons have properties quite different
from those of the quarks.  For instance, they interact incoherently with
external signals.  The elliptic deformation property described in
Fig.~\ref{ellipse} explains the quark and parton models are two
different manifestations of the same covariant entity.

\section{Dirac's coupled oscillators}\label{o32}

Dirac's interest in harmonic oscillators did not stop with his 1945
paper on the representations of the Lorentz group.  In his
1963~\cite{dir63} paper, he constructed a representation of the
$O(3,2)$ deSitter group using two coupled harmonic oscillators.
He starts with two sets of oscillator step-up and step-down operators.
He then ends up with ten operators which act like the generators of
of the $O(3,2)$ deSitter group.  In so doing he constructed the
scientific language of two-photon coherent states or squeezed states
of light which became an important branch of physics 20 years
later~\cite{knp91}.

The $O(3,1)$ Lorentz group is a subgroup of $O(3,2)$.  Therefore,
we are led to suspect that there is a symmetry of Lorentz group in
two coupled harmonic oscillators.  We are particularly interested
in the Lorentz boost property shown in Sec.~\ref{lightcone} and
Fig.~\ref{licone}.

Let us see how these Lorentz-covariant properties are contained
in Dirac's study of the Lorentz group using the two coupled
oscillators.  We start with a simple problem of two oscillators
with equal mass. Then the Hamiltonian takes the form
\begin{equation}
H = {1\over 2}\left\{{1\over m} p^{2}_{1} + {1\over m}p^{2}_{2}
+ A x^{2}_{1} + A x^{2}_{2} + 2C x_{1} x_{2} \right\}.
\end{equation}
This Hamiltonian can be written as
\begin{equation}\label{eq.6}
H = {1\over 2m} \left\{p^{2}_{1} + p^{2}_{2} \right\} +
{K\over 4}\left\{e^{-2\eta}\left(x_1 + x_2\right)^{2} +
e^{2\eta} \left(x_1 - x_2\right)^{2} \right\} ,
\end{equation}
where
\begin{equation}
  K = \sqrt{A^{2} - C^{2}} , \qquad
\exp(2\eta) =\sqrt{\frac{A - C}{A + C} } .
\end{equation}
The wave function then becomes~\cite{hkn99ajp}
\begin{equation}
\psi_{\eta}(x_{1},x_{2}) = {1 \over \sqrt{\pi}}
\exp\left\{-{1\over 4}\left[e^{-\eta}(x_{1} + x_{2})^{2} +
e^{\eta}(x_{1} - x_{2})^{2} \right] \right\} .
\end{equation}
This expression is strikingly similar to the wave function given
in Eq.(\ref{eta}).  It becomes the same if we replace $x_{1}$ and
$x_{2}$ by $z$ and $t$ respectively.

It is indeed remarkable that the Lorentz boost shares the same
geometry as the coupled harmonic oscillators.  It can be seen
from the light-cone view of the Lorentz boost illustrated in
Fig.~\ref{licone}, while the geometry of the coupled
oscillator is basically that of squeezing a circle into ellipse.

\section{One missing component in Dirac's papers}\label{dirac1}

\begin{figure}[thb]
\centerline{\includegraphics[scale=0.7]{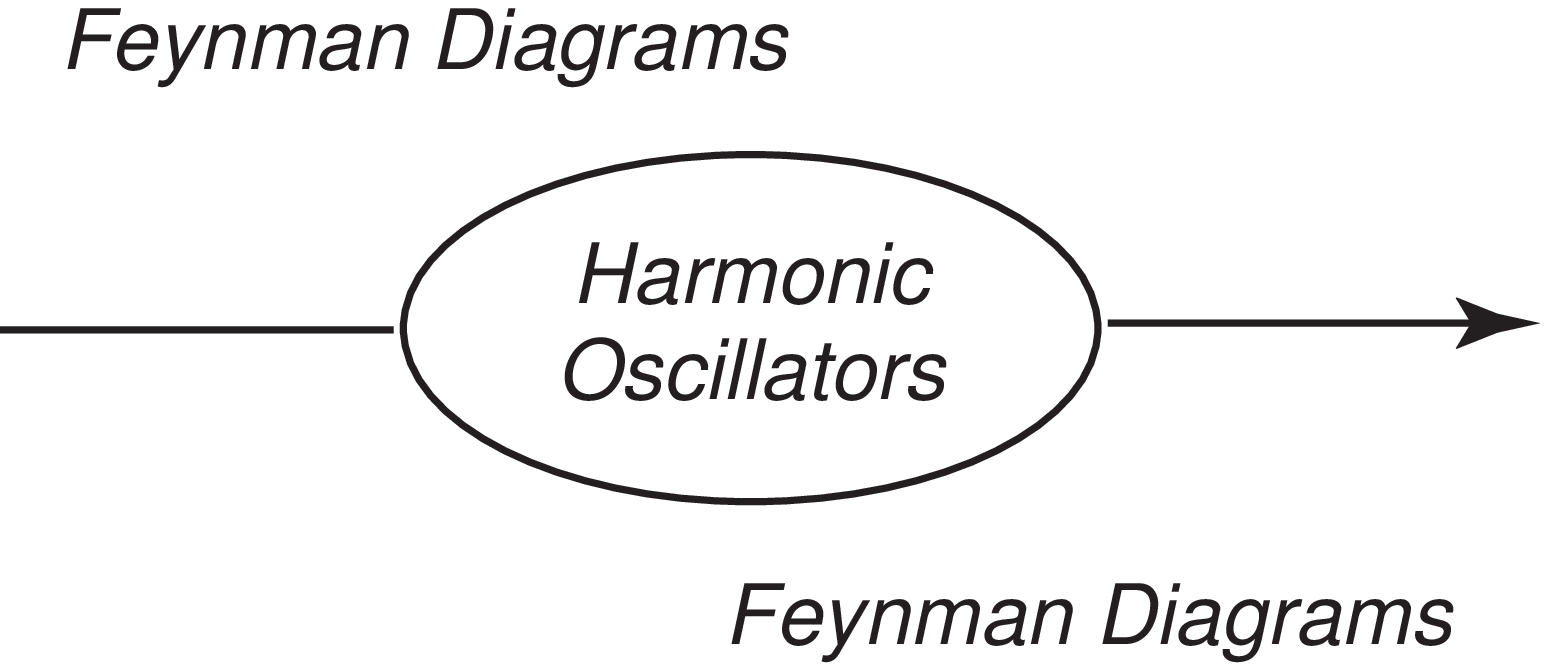}}
\vspace{5mm}
\caption{Feynman's roadmap for combining quantum mechanics with special
relativity.  Feynman diagrams work for running waves, and they provide
a satisfactory resolution for scattering states in Einstein's world.
For standing waves trapped inside an extended hadron, Feynman suggested
harmonic oscillators as the first step.}\label{dff33}
\end{figure}

Quantum field theory has been quite successful in terms of Feynman
diagrams based on the S-matrix formalism, but is useful only for physical
processes where a set of free particles becomes another set of free
particles after interaction.  Quantum field theory does not address the
question of localized probability distributions and their covariance
under Lorentz transformations.  In order to address this question,
Feynman {\it et al.} suggested harmonic oscillators to tackle the
problem~\cite{fkr71}.  Their idea is indicated in Fig.~\ref{dff33},
and also in Fig.~\ref{comet}.

In this report, we are concerned with quantum bound system, and we
have examined the four-papers of Dirac on the question of making
the uncertainty relations consistent with special relativity.  Indeed,
Dirac discussed this fundamental problem with mathematical devices
which are both elegant and transparent.

Dirac of course noted that the time variable plays the essential
role in the Lorentz-covariant world.  On the other hand, he did
not take into consideration the concept of time separation.
When we talk about the hydrogen atom, we are concerned with the
distance between the proton and electron.  To a moving observer,
there is also a time-separation between the two particles.

\begin{figure}[thb]
\centerline{\includegraphics[scale=0.8]{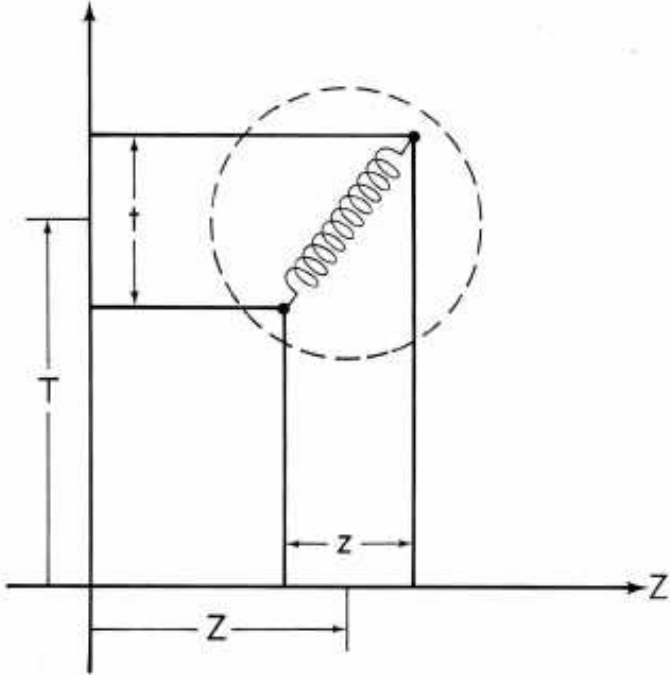}}
\caption{Space and time separations in the Lorentz-covariant world.
Wherever there is a space-separation, there is a time-separation.
Two simultaneous events separated by a distance are not simultaneous
for moving observers.}\label{tsep}
\end{figure}

Instead of the hydrogen atom, we use these days the hadron
consisting of two quarks bound together with an attractive force,
and consider their space-time positions $x_{a}$ and $x_{b}$, and
use the variables~\cite{fkr71}
\begin{equation}
X = (x_{a} + x_{b})/2 , \qquad x = (x_{a} - x_{b})/2\sqrt{2} .
\end{equation}
The four-vector $X$ specifies where the hadron is located in space
and time, while the variable $x$ measures the space-time separation
between the quarks.  Let us call their time components $T$ and $t$
as illustrated in Fig.~\ref{tsep}.  These variables actively
participate in Lorentz transformations.  The existence of the
$T$ variable is known, but the Copenhagen school was not able to
see the existence of this $t$ variable.

Paul A. M. Dirac was concerned with time variable throughout
his four papers discussed in this report.  However, he did not make
a distinction between the $T$ and $t$ variables.
The $T$ variable ranges from $-\infty$ to $+\infty$, and is constantly
increasing.  On the other hand, the $t$ variable is the time
interval, and remains unchanged in a given Lorentz frame.

Indeed, when Feynman {\it et al.} wrote down the Lorentz-invariant
differential equation~\cite{fkr71}
\begin{equation}\label{osceq}
{1\over 2} \left\{x^{2}_{\mu} -
{\partial^{2} \over \partial x_{\mu }^{2}}
\right\} \psi(x) = \lambda \psi(x) ,
\end{equation}
$x_{\mu}$ was for the space-time separation between the quarks.

This four-dimensional differential equation has more than 200
forms of solutions depending on boundary conditions.  However,
there is only one set of solutions to which we can give a physical
interpretation.  Indeed, the Gaussian form of Eq.(\ref{ground4})
is a solution of above differential equation.  If we boost the
system along the $z$ direction, we can separate away the $x$ and
$y$ components in the Gaussian form and write the wave function
in the form of Eq.(\ref{ground}).

It is then possible to construct a representation of the
Poincar\'e group from the solutions of the above differential
equation~\cite{knp86}.  If the system is boosted, the wave function
becomes the Gaussian form given in Eq.(\ref{eta}), which becomes
Eq.(\ref{ground}) if $\eta$ becomes zero.  This wave function is
also a solution of the Lorentz-invariant differential equation of
Eq.(\ref{osceq}).  The transition from Eq.(\ref{ground}) to
Eq.(\ref{eta}) is illustrated in Fig.~\ref{ellipse}.

\section*{Concluding Remarks}
The easiest way to build a canal is to link up existing lakes.
Paul A. M. Dirac indeed dug four big lakes.  It is a pleasure to
link them up.  Dirac constructed those lakes in order to study
whether the Copenhagen school of quantum mechanics can be made
consistent with Einstein's Lorentz-covariant world.

After studying Dirac's papers, we arrived at the conclusion that
the Copenhagen school completely forgot to take into account
the question of simultaneity and time separation~\cite{kn06aip}.
The question then is whether the localized probabilty distribution
can be made consistent with Einstein's Lorentz covariance.

We would like to thank T. R. Love for helpful comments and for
pointing out a number of typographical  errors in the preliminary
version of this paper.

\end{document}